\documentclass[12pt,preprint]{aastex}
\usepackage{psfig}
\usepackage{emulateapj5}

\def\msun{M$_{\odot}$}
\def\eg{{\em eg.\ }}

\def\etal{{\em et al.\ }}
\def\ie{{\em i.e.\ }}
\def\lesssim{\mathrel{\hbox{\rlap{\hbox{\lower4pt\hbox{$\sim$}}}\hbox{$<$}}}}
\def\gtrsim{\mathrel{\hbox{\rlap{\hbox{\lower4pt\hbox{$\sim$}}}\hbox{$>$}}}}

\slugcomment{Accepted for publication in AJ}

\shorttitle{}
\shortauthors{}

\begin{document}


\title{Evidence for Stellar Substructure in the Halo and Outer Disk of M31 \footnotemark[1]}
\footnotetext[1]{Based on observations made with the Isaac NewtonTelescope operated on the Island of La Palma by the Isaac Newton Group
in the Spanish Observatorio del Roque de los Muchachos of the Instituto
de Astrofisica de Canarias}
 

\author{Annette M.~N. Ferguson}
\affil{Kapteyn Astronomical Institute, Postbus 800, 9700 AV Groningen, The
       Netherlands}
\email{ferguson@astro.rug.nl}
\author{M.~J. Irwin}
\affil{Institute of Astronomy, Madingley Road, Cambridge UK CB3 0HA}
\email{mike@ast.cam.ac.uk}
\author{R.~A. Ibata}
\affil{Observatoire de Strasbourg, 11, rue de l'Universit\'{e}, F-67000, Strasbourg, France}
\email{ibata@newb6.u-strasbg.fr}
\author{G.~F. Lewis}
\affil{Anglo-Australian Observatory, P.O. Box 296, Epping, NSW 1710, Australia}
\email{gfl@aaoepp.aao.gov.au}
\author{N.~R. Tanvir}
\affil{Department of Physical Science, University of Hertfordshire, College Lane, Hatfield UK AL10 9AB}
\email{nrt@herts.star.ac.uk}


\begin{abstract}
We report the discovery of significant stellar substructure in the halo
and outer disk of our nearest large galactic neighbour, M31.  Our deep
panoramic survey with the Isaac Newton Telescope {\it Wide Field
Camera} currently maps out an area of $\approx 25~\sq\arcdeg$ around
M31, extending along the semi-major axis to 55~kpc, and is the first to
allow an uninterrupted study of the density and color distribution of
individual red giant branch stars across a large fraction of the halo
of an external spiral galaxy.  We find evidence for both spatial
density and metallicity (as inferred from colour information)
variations, which are often, but not always, correlated.  In addition
to the previously reported giant stellar stream \citep{ibata01b}, the
data reveal the presence of significant stellar overdensities at large
radii close to the south-western major axis, in the proximity of the
very luminous globular cluster G1, and near the north-eastern major
axis, coinciding with and extending beyond the previously-known
$`$northern spur'.  The most prominent metallicity variations are found
in the southern half of the halo, where two large structures with above
average metallicites are apparent; one of these coincides with the
giant stellar stream while the other corresponds to a much lower-level
stellar enhancement.  Our findings contrast with, but do not conflict
with, past studies of the M31 halo and outer disk which have suggested
a rather homogeneous stellar population at large radius: the bulk of
our newly-detected substructure lies in the previously-uncharted far
outer regions of the galaxy.  We discuss the possible origin of the
substructure observed and the implications it has for constraining the
galaxy assembly process.
\end{abstract}

\keywords{galaxies: individual (M31)-- galaxies: evolution -- galaxies: halos --
Local Group -- galaxies: stellar content -- galaxies: structure}

\section{Introduction}

Within the context of the hierarchical clustering theory for structure
formation, large disk galaxies like the Milky Way and M31 arise from
the merger and accretion of many smaller subsystems and from the smooth
accretion of intergalactic gas (\eg \citet{white78, stein02}).  While
the former process governs the growth of the dark matter halo (by far
the dominant mass component in a disk galaxy) and much (if not all) of
the stellar spheroid, angular momentum-conserving collapse of the
latter is the currently-favoured mechanism for forming the thin disk.
It has  been argued that the relative importance of these processes --
\ie  smooth  versus discrete accretion --  in the assembly of a
galaxy is the primary factor which determines the final morphology (\eg
\citet{baugh96,stein02}).

Cosmological simulations of galaxy formation have become increasingly
sophisticated in recent years and have now attained sufficient
resolution to begin to address the internal structures of galaxies.  A
key result to emerge from state-of-the-art simulations within the
popular cold dark matter (CDM) framework is that the dark matter
subhalos which merge to form massive galaxies are much more resilient
than previously thought \citep{klypin99,moore99}.  Although dynamical
friction and galactic tidal forces continually act to disrupt subhalos
once they fall within the potential of the massive system, the dense
central cores of the satellites appear to survive as distinct entities
for at least several orbital timescales (\eg \citet{hay02}).  The
total mass in undisrupted cores is expected to be quite small
($\lesssim$ 10\% of the total mass of the system) with the bulk of the
mass distributed much more smoothly, however several hundred of them
are expected to reside within the virial radius of a typical galaxy
like the Milky Way.

Signatures of the hierarchical nature of galaxy assembly are expected
to be most obvious in the properties of the halo stellar populations.
At least some of the accreted satellites will have experienced
significant star formation prior to and during incorporation into the
final system.  As these systems will span a range of masses and have
experienced different dynamical histories within the massive galaxy
potential, it seems highly likely that they will also be characterised
by distinct star formation and chemical enrichment histories, the
details of which will be imprinted on the ages and metallicities of
their constituent stars.  Indeed, studies of Milky Way satellite
stellar populations reveal a surprising variety of complex star
formation histories, with no two dwarfs sharing a similar evolution
\citep{mateo98b}.  During the accretion of a satellite galaxy onto a
massive host, tidal forces first shock the dark matter, then
the stars, of the satellite making them gravitationally
unbound.  Tidal debris will be deposited in both a leading and a
trailing stream which, depending on the satellite's orbit, initial
mass and sphericity of the host potential, can maintain spatial and
kinematic coherence within the halo for many Gyr \citep{john96,
helmi99a,white00,bull01}.  It is therefore expected that the stellar
halos of galaxies which form hierarchically should posses significant
spatial and metallicity substructure in the form of disrupted
satellites (\eg dwarf galaxies, globular clusters) and their stellar
detritus.  This structure will be particularly apparent in the outer
regions of the halo, where the dynamical mixing timescales are long.
Furthermore, as there is no reason to believe that the accretion
history will be exactly the same in different galaxies (for example, the number
and masses of accreted satellites and their mix of stellar
populations), one might expect to see noticeable differences in the
properties of galactic stellar halos, even between systems of similar
morphological type.

It has also been postulated that some fraction of the accreted
satellites lack a luminous stellar component, either due to feedback
from early star formation which expels the gas before
significant star formation can occur (\eg \citet{white78,ds86}) or
from the inhibition of gas cooling due to an ionizing UV background
(\eg \citet{bull00}).  The main reason for invoking these scenarios is
the striking discrepancy between the numbers of satellites predicted
to lie within the dark halo of a massive disk galaxy and the numbers
of dwarf companions actually observed around the Milky Way and M31
\citep{klypin99,moore99}.  If correct, detecting these CDM
substructures through stellar density enhancements becomes very
difficult, if not impossible.  On the other hand, such clumpy
structures will still exert an influence on the the host galaxy by
causing tidal heating and/or distortion of the thin stellar disk,
particularly in the fragile outer regions \citep{moore99,font01}.

 Searching for and studying stellar substructure in the outskirts of
disk galaxies provides an important test of CDM models of
structure formation.  Studies of the Milky Way stellar halo have so
far provided encouraging support for hierarchical galaxy assembly.
Evidence includes the discovery of the tidally-disrupted Sagittarius
dwarf galaxy and associated stellar streams (\eg
\citet{ibata95,mateo98a,maj99,yanny00,ibata01a,ibata01c}), the
phase-space clumping of halo stars in the solar neighbourhood
\citep{helmi99b} and the tidal tails emanating from halo globular
clusters (\eg \citet{grillmair95,oden01,leon00}) and dwarf Spheroidal
galaxies (\eg \citet{irwin95,maj00}).  Unfortunately, our location
within the disk sometimes renders interpretation of structures revealed
through star counts rather difficult (\eg Newberg \etal 2002)
and underscores the need for an extragalactic perspective.  It is thus 
clearly desirable to test whether the halos of other disk galaxies
also exhibit substructure and to investigate how the nature of this
substructure (as well the nature of the field population resulting
from totally disrupted satellites) varies with the properties of the
host galaxy.

 Searches for low surface brightness tidal and/or extraplanar features
have been carried out around several external edge-on disk systems but
to date only one unambiguous detection has been made, in NGC~5907
\citep{shang98}.  On the other hand, such studies are technically very
difficult, and inaccurate flat-fielding and/or low-level scattered
light can easily mask real signal at these levels ($\Sigma_V \gtrsim
28$ magnitudes arcsec$^{-2}$). A more preferable search technique
involves using resolved star counts to map out the spatial density
structure in nearby galaxy halos, however, with ground-based
telescopes, this method is currently limited to galaxies within and
around the Local Group. As these galaxies also subtend the largest
angular sizes on the sky, quantitative study requires wide-field CCD
mosaic cameras as well as sophisticated processing techiques capable of
dealing in an optimal way with large amounts of data, both of which
have only become available in recent years.

Using the {\it Wide-Field Camera} on the Issac Newton Telescope (INT
WFC), we are carrying out a panoramic imaging survey of our nearest
large neighbour, M31.  The contiguous nature of our survey allows us
to distinguish local density enhancements in M31's halo and outer disk
from fluctuations in background galaxy counts and the foreground
distribution of Galactic stars.  The present paper reports interim
results from this survey, namely the discovery of significant spatial
density and metallicity substructure in the red giant population of
the halo and outer disk.  The stellar halo of M31 has long been known
to have rather different properties from the Milky Way halo, despite
the overall similarity in the global properties of the galaxies.  In
particular, ground-based and HST studies indicate the M31 halo is
roughly an order of magnitude more metal-rich (\eg
\citet{morris94,rich96,holland96,durrell01,reitzel02}) and denser
and/or larger \citep{reitzel98} than that of the Milky Way.  Our INT
WFC survey reveals significant stellar substructure in the outskirts of
M31 and provides the tantalizing suggestion that these halo differences
may be due to a more active accretion/merger history within the M31
subsystem.  Preliminary results, drawn from the first phase of our
survey, were reported in \citet{ibata01b}.

\section{Observations}

The {\it Wide Field Camera} on the 2.5m Isaac Newton Telescope is a
4-chip EEV 4k$\times$2k CCD mosaic camera which images $\approx$0.29
square degrees per exposure \citep{walton}.  On the nights of 3-9 of
September 2000, 9-16 October 2001 and 13 November 2001, we used this
camera to image 91 contiguous fields (corresponding to
$\approx~25~\sq\arcdeg$) in the outer disk and halo of M31.  Coverage
currently extends to 4.0\arcdeg ($\approx 54$~kpc) and 2.5\arcdeg
($\approx 34~$kpc in projection) along the major and minor axes
respectively.

Images were taken in the equivalent of Johnson V and Gunn $i$ bands
under mainly good atmospheric conditions, with 85\% of the fields taken
in photometric conditions with seeing better than 1.2\arcsec.  The
exposure time of 800-1000s per passband per field allowed us to reach
$i=23.5$, V$=24.5$ (S/N$\approx$5) and is sufficient to detect
individual red giant branch (RGB) stars to M$_V \approx 0$ and main
sequence stars to M$_V \approx -1$ at the distance of M31.  Several
fields taken in poorer conditions were reobserved and coadded as
necessary to give an approximately uniform overall survey depth.

All of the on-target data plus calibration frames were processed
 using the standard INT Wide Field Survey pipeline provided by the
Cambridge Astronomical Survey Unit \citep{irwin01}.  This package
provides the usual facilities for instrumental signature removal,
including in this case defringing of the $i$-band data, plus tools for
object catalogue creation, astrometric and photometric calibration,
morphological classification and cross-matching catalogues from
different observations.  The pipeline processing provides internal
cross-calibration for the four CCDs at a level better than 1\% within
each pointing.  Field-to-field variations in photometric zero-points
were calibrated and cross-checked using a combination of multiple
nightly photometric standard sequence observations and the overlap
regions between adjacent WFC pointings.  The overall derived
photometric zero-points for the whole survey are good to the level of
$\pm 2$\% in both bands.  The full details of the survey strategy, data
processing and calibration will be presented elsewhere (Irwin \etal, in
prep).

Objects were classified as noise artifacts, galaxies, or stars
according to their morphological structure on all the images.  As a
sanity check, we compare our extended source counts in the outer halo
fields with the deep V and I band galaxy counts of \citet{smail95}.
Over the magnitude ranges $22.5<$V$<23.5$ and $20.5<i<21.5$, we detect
$\approx 11500$ and $\approx 7300$ galaxies per square degree
respectively, which can be compared with the $\approx 10000$ per square
degree in each band predicted using \citet{smail95} and indicates that
we are not mis-classifying large numbers of faint galaxies.  In the
outer halo fields, we typically detect equal numbers of stars and
galaxies within the magnitude and color ranges of interest (see Section
3.1 and Figure \ref{sel}).  Of the extended sources, approximately 20\%
are $`$compact' in the sense of being within the 3-5 sigma range of the
stellar boundary in the classification statistic and having an
ellipticity of $<0.4$.  Making the plausible assumption that half of
these are genuine galaxies (the other half being  genuinely stellar)
leads us to we expect that the contamination due to mis-classified
barely resolved field galaxies is small and generally considerably
less than 10\% of the total number of detected sources.  Indeed, the
overdense regions we will discuss in this paper have stellar densities
at least a factor of two higher, and the average star counts are a
factor of three higher, than that in the outer halo fields.

Stellar contamination from the Galactic foreground steadily
increases to the northeast, due to the proximity of the Galactic Plane,
rising smoothly from an average contamination of $\approx$13000 stars
per square degree at the southwest extremity of the survey to
$\approx$20000 stars per square degree at the northeast extremity
(integrated over all magnitudes).  This foreground variation, coupled
with our current lack of suitable comparison fields uncontaminated by
M31, constrains to some extent detailed quantitative analysis of the
spatial and metallicity distribution of the outer parts of the halo.

\section{Results}

\subsection{Spatial Density Variations}

\begin{figure}
\centerline{\psfig{file=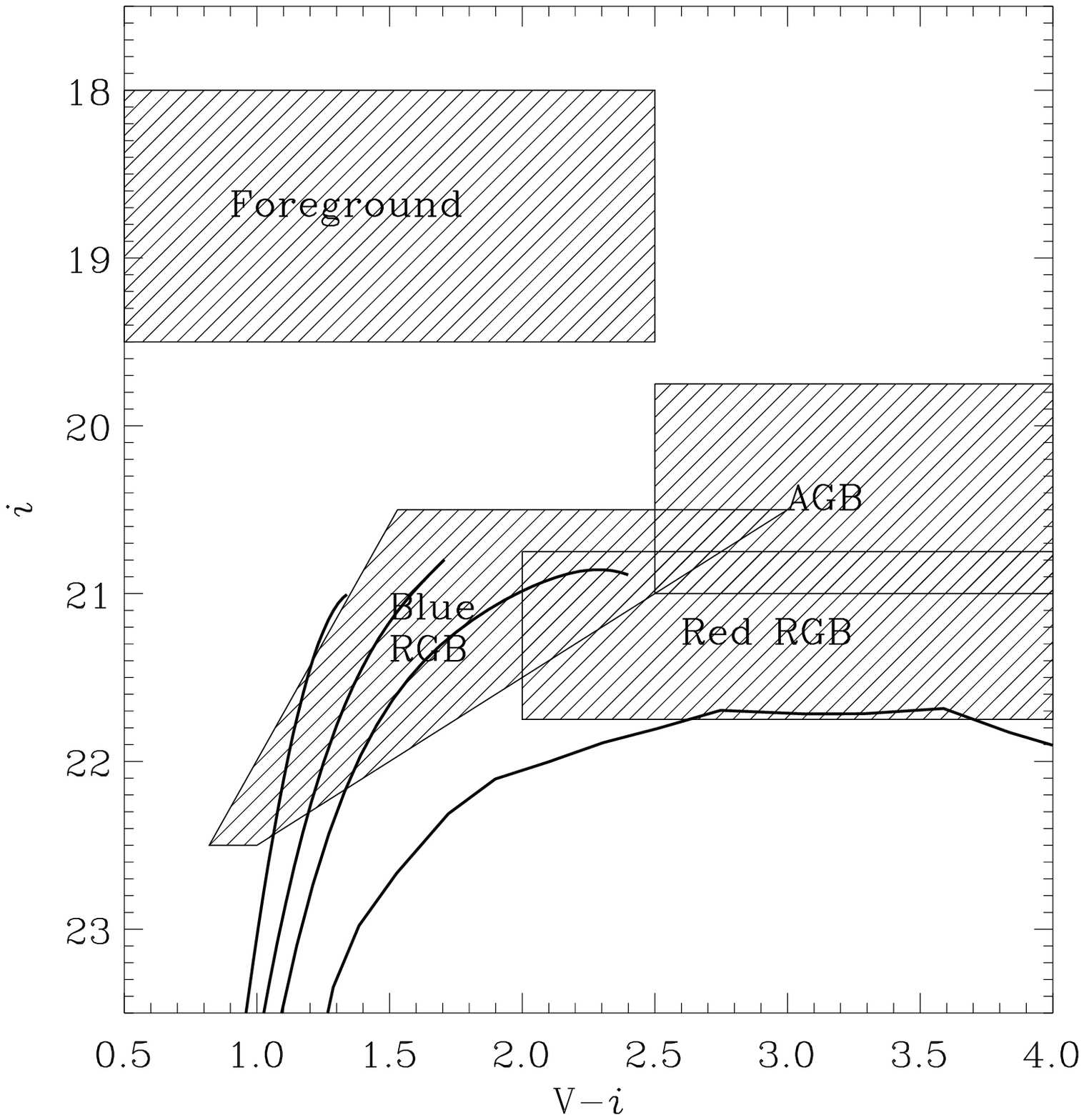,height=15cm}}
\caption{An illustration of the selection criteria adopted to isolate
stars in different regions of the color-magnitude diagram.  The blue
RGB ($20.5<i<22.5; 8.7-0.4~i<V-i<23.5-i$) and red RGB ($20.75<i<21.75;
V-i>2.0$) cuts are designed to select stars from the metal-poor and
metal-rich sides of the red giant branch respectively. The AGB cut
($19.75<i<21.00;V-i>2.5$) is designed to select intermediate-age,
moderate metallicity stars in the thermally-pulsing regime of the
asymptotic giant branch. Field-to-field corrections for the varying
Galactic foreground population (see Section 3.2) are calculated using
star counts in the region defined by $18.0<i<19.5$ and $0.5<V-i<2.5$.
Also overlaid are the giant branch fiducial sequences of several
Galactic globular clusters: from left to right these correspond to
NGC~6397 ([Fe/H]$=-1.9$), NGC~1851 ([Fe/H]$=-1.3$), 47~Tuc
([Fe/H]$=-0.7$) and NGC~6553 ([Fe/H]$=-0.3$) taken from
\citet{dacosta90} and \citet{sagar99} (see text).  In placing the
fiducials on the (V,$i$) plane, we have made use of the color equations
derived for the WFS data (see
http://www.ast.cam.ac.uk/$\sim$wfcsur/photom.html).\label{sel}}
\end{figure}

Our INT WFC survey of M31 provides the first opportunity to make an
uninterrupted study of the properties of the resolved stars across a
large fraction of an external disk galaxy.  $`$Stellar' sources in our
catalogue consist of M31 red giants, M31 upper main sequence stars,
Galactic foreground stars and unresolved background galaxies.  In order
to examine the spatial distribution of various stellar populations
within M31, we apply a series of magnitude and color cuts which are
designed to isolate stars in different regions of the color-magnitude
diagram (CMD).  Figure \ref{sel} illustrates the main selection
criteria adopted in our analysis.  We define the $`$blue RGB' as stars
with $20.5<i<22.5; 8.7-0.4~i<V-i<23.5-i$ and the $`$red RGB' as stars
with $20.75<i<21.75;V-i>2.0$.  These magnitude and color ranges are
consistent with those expected for red giant stars at the distance of
M31.  We also overlay several Galactic globular cluster fiducials
spanning a range in metallicity. Comparison between our selection
criteria and these fiducials demonstates that our RGB cuts isolate the
metal-poor ($-2.0\lesssim$[Fe/H]$\lesssim -0.6$)  and metal-rich
($-0.6\lesssim$[Fe/H]$\lesssim -0.3$) sections of the giant branch
respectively.

Figure \ref{rgb} shows the standard coordinate projection of the
surface density distribution of blue and red RGB stars across our
current $\approx 25~\sq \arcdeg$ survey area. Each detected source
classified as stellar on the $i$-band and falling within the
aforementioned magnitude and color limits is encoded as a point on
these diagrams.  The lower magnitude limits are conservatively set at
roughly one magnitude brighter than the survey 5-$\sigma$ detection
threshold to mitigate the effects of varying completeness.  The
immediate impression from Figure \ref{rgb} is the striking
non-uniformity of the stellar distribution at large radii.  The
large-scale morphology of stars in the blue RGB map is that of a
significantly flattened inner halo structure, while the bulk of the
stars in the red RGB map appear associated with the outskirts of the
stellar disk.   Numerous stellar enhancements are present in both maps,
with some features appearing more conspicuous in one color cut than
another.    The giant stellar stream discovered during the first phase
of our survey \citep{ibata01b} is apparent as an enhancement close to,
but distinct from, the southern minor axis and extending out to
$\approx 40$~kpc.  While visible in both maps, the feature appears
considerably sharper in the distribution of red RGB stars.  In
addition, significant stellar overdensities are seen at large radii
lying close to both major axes (at 1.5\arcdeg~W, 1.8\arcdeg~S and
0.8\arcdeg~E, 1.8\arcdeg~N). A lower intensity enhancement is visible
in the form of a diffuse extension northeast of the center (towards
1.7\arcdeg~E, 0.7\arcdeg~N).    Very faint structure, some of which is
at the limit of detectability, is seen emanating from the northern side
of the disk.    No correction has been applied to the maps for Galactic
foreground contamination, however we have verified that the
distribution of stars with properties expected to belong to this
population (selected from the box indicated in  Figure \ref{sel}) is
smooth across the entire survey area.  It is therefore unlikely that
Galactic foreground variations are causing any of the structure seen in
Figure \ref{rgb}.  Likewise, the Galactic extinction towards the
outskirts of M31 is relatively uniform, ranging from
E(V$-i$)=0.06--0.11 \citep{schl98}.  There is no evidence for
significant clumping in the distribution of extended sources (\ie
background galaxies) across our survey area and, in any case,  galaxy
clusters and other large scale structure would be unlikely to dominate
the counts at the magnitudes, colors and angular scales of the observed
features.  Indeed, \cite{couch93} show that the galaxy-galaxy
correlation function is very low beyond 0.1\arcdeg\ at the magnitudes
of relevance for our survey.    We thus conclude that the spatial
density enhancements reflect real substructure associated with M31's
halo and outer disk.

The substructure located near the southwestern major axis is of
particular interest because it is located in the proximity of the very
luminous globular cluster G1 \citep{meylan01}. This clump, which we
will hereafter refer to as the $`$G1 clump' although the nature of the
association with G1, if any, is presently unclear, is located at $\sim
35$~kpc in projected radius and has a physical size of $\sim 0.7\arcdeg
\times 0.5\arcdeg$, or $10\times7$~kpc at the distance of M31. Compared
with the mean density of RGB stars at similar radii around M31, we
calculate that the G1 clump is overdense by a factor of $\approx$4.
Integrating the excess RGB population over this region relative to
nearby comparison fields and assuming a luminosity function similar to
the field population studied by \citep{rich96} gives estimates for the
total apparent magnitude of this feature of $m_V = 12.1$ and $m_i =
10.5$, with average surface brightnesses of around 28.5 and 27.0
magnitudes arcsec$^{-2}$ respectively. At an average extinction of
E(B-V) = 0.07 this is equivalent to absolute magnitudes of  $M_V =
-12.6$ and $M_i = -14.1$.  While the magnitude and color of the clump
are comparable, within the errors, to those of dwarf satellites in the
M31 and Milky Way sub-groups, the measured surface brightness is
several magnitudes fainter than any of the currently known systems; on
the other hand, it is similar to that measured for the low surface
brightness ring in NGC~5907 \citep{shang98}.
 
The density enhancement lying near the northeastern axis coincides with
and extends beyond the $`$northern spur', first remarked upon by
\citet{wk88} (hereafter WK88) as faint light bending away from the
major axis (see also the deep images of \citet{inn82}). WK88 questioned
whether this feature was actually associated with M31, or due to a
Galactic reflection nebula: our color-magnitude diagrams of this
region unambiguously place the excess stars at the same distance as
M31.  The feature extends to a projected radius of $\sim30$~kpc at
angles up to $\sim20\arcdeg$ off the major axis.  The northern spur
region is about a factor of 1.5 -- 2 times more overdense than the G1
clump and thus has a mean surface brightness
$\approx0.5-0.8$~magnitudes arcsec$^{-2}$ higher.  The northern spur
region also appears as a low level enhancement on a spatial density
plot of stars selected to lie just above and redward of the RGB tip
(see Figures \ref{sel} and \ref{agb}). Such stars are likely to be
intermediate-age, moderate-metallicity ([Fe/H] $\gtrsim -0.7$; ages
$\gtrsim 3-8$~Gyr) thermally-pulsing asymptotic giant branch (AGB)
stars \citep{girardi00}, and their presence is taken as evidence for an
extended epoch of star formation in this region.   There is also a
very tentative detection of AGB stars in the vicinity of the G1 clump,
however all other substructure appears devoid of this younger
population.  While the effects of crowding will lead to some spurious
structure on the AGB map (\eg the strong detection of NGC~205), this is
unlikely to be important in the diffuse outer regions of the galaxy
under study here.

Perhaps surprisingly, our current map of the RGB density around M31
reveals no obvious northern counterpart to the giant stellar stream
discovered near the southern minor axis \citep{ibata01b}. We will
return to this issue in the discussion below.

\subsection{Metallicity Variations}

Further insight on the nature of the stellar density variations in
M31's halo and outer disk comes from analysis of the spatial variation
of mean RGB color. Figure \ref{rgb} already reveals that the morphology
of the M31 substructure has a color dependence.  This is further
illustrated in Figure \ref{cmds}, which shows color-magnitude diagrams
for fields in the giant stellar stream (right) and the G1 clump
(left).  Comparison with Galactic globular cluster fiducials reveals
that the mean RGB color varies between these pointings, being slightly
more metal-rich than 47~Tuc in the stellar stream and approximately
equal to 47~Tuc in the G1 clump.  The width of the G1 clump RGB also
appears somewhat narrower than that of the stellar stream, indicating a
small intrinsic metallicity and/or age dispersion.  In order to
quantify the RGB color variation as a function of position over the
survey area, it is first necessary to make a statistical correction for
the foreground Galactic contamination.  Three outer halo fields were
used to define a reference foreground population (albeit with a small
halo contamination still present in these fields) and the scale factor
required to match the target field $i$-band luminosity function
brighter than $i = 20$ computed using the foreground box in Figure
\ref{sel}.  For each field, we then calculate the foreground-corrected
median color distribution in the magnitude range $21 < i < 22$
projected in a coordinate system orthogonal to the locus of the
ensemble RGB for M31.  This color measure quantifies the shift in the
mean RGB locus as a function of field pointing. Figure \ref{metal}
shows a color-coded plot of this variation. The $`$average' halo RGB
color is denoted here as green; assuming an old stellar population,
this corresponds to a metallicity of [Fe/H]$\sim -0.7$, similar to
47~Tuc. Progressively bluer shades indicate bluer mean RGB colors while
progressively redder shades indicate redder mean RGB colors.

We adopt the view here that metallicity variations are the dominant
driver of these color variations.  While age variations may contribute
somewhat to the observed behaviour, there is yet to emerge any
compelling evidence for a luminous AGB component in M31's extended halo
which would accompany a young-to-intermediate age population (\eg
\citet{rich96, holland96}; also see Figure \ref{agb}).   For
reference,  an age variation of $\sim15$~Gyr at $[$Fe/H$]=-0.7$ would
be required to explain the entire color variation seen over the survey
area \citep{girardi00}. Based on the assumption of a uniformly old age
for the M31 halo, comparison with Galactic globular cluster tracks and
theoretical isochrones indicates that color index variations of $\pm
0.2$ in V$-i$ (the full range of colors used in Figure \ref{metal})
correspond to a metallicity variations ranging from -0.5 dex below the
mean to 0.25 dex above the mean.  As the relationship between RGB color
and metallicity is highly non-linear across the range of interest (see
Figure \ref{cmds}), the quoted spread in metallicities is unfortunately
rather uncertain.  The observed color variations are well in excess of
our photometric errors at these magnitudes, and also significantly
larger than the zeropoint uncertainties.  Although our strategy of
bootstrapping the calibration of non-photometric pointings from
adjacent fields could introduce some correlated color variations, the
size of this effect is also estimated to be rather small, amounting to
no more than a few hundredths of a magnitude.

Inspection of Figure \ref{metal} reveals that while there are no
obvious radial metallicity gradients present in the halo, there are
significant chemical inhomogeneities on large (\ie several kpc)
scales.  Further, comparison of Figure \ref{metal} and Figure
\ref{rgb} indicates that while substructure in the
metallicity map often corresponds to overdensities on the spatial
density map, this is not universally true.    The most prominent
chemical substructure is found on the southern side of the galaxy, in
the region of the stellar stream and further to the north on the same
side, where only a diffuse extension of the halo is present on the
spatial density map. Both these regions appear significantly more metal
rich than the average halo and it is tempting to speculate that the two
might be related, possibly tracing out the projected orbit of the
object producing the stellar stream.  We note that line-of-sight depth
through the halo may cause intrinsic metallicity variations to be
somewhat $`$washed out', depending on how localised the substructure is in
space and how large the overdensity is. The true metallicity variations
in the halo are thus likely to be somewhat underestimated in
Figure \ref{metal}.

Another apparently metal-rich region is the northern spur; at first
glance this might appear to be even more metal rich than the outer
stellar disk (the edge of which is approximately delineated by the
inner ellipse) however the disk fields contain OB stars which make them
appear bluer, leading to spuriously low metallicity estimates.
Similarily, age effects also complicate the interpretation of the spur,
where we have evidence for the presence of an intermediate-age
population (Figure \ref{agb}).  The metal-rich nature of this feature
appears rather robust though, given that the young population will
cause the instrinsic stellar colors to appear bluer, not redder.

The G1 clump does not appear particularly distinct on the metallicity
map, contrary to its appearance on the spatial density map, with the
RGB stars in this region appearing just slightly more metal poor than
the background halo.

\section{Discussion}  
\subsection{Comparison to Previous Studies of the M31 Outer Disk and Halo}
\subsubsection{Optical Studies}

Many previous studies have addressed the stellar population
characteristics of the halo and outer disk of M31, however these have
almost exclusively sampled only a few discrete locations or have taken
a panoramic, but much shallower view.  Early quantitative wide area
survey work of M31 was based on photographic plates (\eg
\citet{hodge73,inn82,wk88}).  Although these studies clearly revealed,
among other things, the anti-symmetric warping of the outer disk and
the tidally-induced twisting of the outer isophotes of the dwarf
companions (see also \citet{choi02}), they did not go deep enough to
directly resolve the stellar populations.

More recent work has focused on detailed studies of resolved stars in
several (generally small) fields sampling the outer disk and halo (see
for example
\citet{mould86,morris94,rich96,holland96,reitzel98,durrell01,ferg01,sara01,reitzel02}).
Both ground-based and HST studies have found a dominant population
which is significantly more metal-rich than that of the Milky Way halo
and which is characterised by a considerable intrinsic dispersion.
Specifically, measurements indicate $[$Fe/H$] \sim -0.7$, comparable to
47~Tuc, with spread of nearly 2 dex for M31, compared to $[$Fe/H$] \sim
-1.5$ for the Milky Way.  Detailed metallicity distributions have now
been calculated for stars in several fields, all of which are found to
exhibit a similar shape, consisting of a peak at the metal-rich end and
an extended tail towards lower metallicities
\citep{holland96,durrell01,sara01,reitzel02}.  The fraction of stars in
the metal-poor component is estimated to be 25-50\% with no compelling
evidence for a strong metallicity gradient and/or inhomogeneities in
the halo. Additionally, \citet{reitzel98} find that M31's halo is
roughly an order of magnitude larger and/or denser than the Galactic
halo, though this conclusion is drawn from star counts in only three
fields.

Analysis of individual color-magnitude diagrams from our WFC survey
indicates a mean metallicity in broad agreement with the findings of
previous studies (see Figure \ref{cmds}; also \citet{ibata01b}) and
reveals no obvious metallicity gradient in the halo out to at least
50~kpc.  We do, however, find evidence for significant spatial and
metallicity substructure in the halo.  Figure \ref{prev} shows the
location of published HST fields and a few recent ground-based fields
overlayed on our survey area.  It can be seen that these fields
generally sample either the inner halo of M31 or at large radius along
the southern minor axis, neither of which exhibit significant
substructure in our survey.  The chemical and spatial homogeneity of
the halo that has been inferred from these studies is therefore not
surprising.  Most of the substructure our survey has uncovered lies in
the previously-uncharted outer regions of M31.

An exception to this is the study by \citet{morris94} which resolved
the stars in a small section of the northern spur and provided the
first evidence that the stellar overdensity in these parts consisted of
the fairly characteristic metal-rich population seen elsewhere in the
outskirts of M31.  Like us, these authors also found evidence for a
significant AGB component lying above the RGB tip.  In addition,
\citet{rich96} obtained an HST/WFPC2 CMD of the field population around
the globular cluster G1, corresponding to a location near the very edge
of the stellar overdensity discovered in our WFC survey, and found the
RGB to be well-matched by the 47~Tuc fiducial. They also noted that the
luminosity function of the  field population could not be fitted by
template globular cluster luminosity functions.   While the results
from these studies are in good qualitative agreement with those
presented here,  their lack of depth and/or areal coverage adds little
further insight into the nature of the stellar populations associated
with the substructure.

\subsubsection{HI Studies}

The most extensive published wide-field survey of HI in M31 is still
that of \citet{newt77} who mapped regions along the major axes out to
$\sim 35$~kpc.  Apart from the warp in the disk at large radius, the
outermost  HI contour looks remarkably smooth and uniform with no sign
of perturbation in the direction of any of the halo substructure (see
their Figure 12a).  Furthermore, while the optical and HI disks are
warped in the same direction, the optical warp appears to begin at
smaller radii than that of the HI and exhibit a greater deviation from
the plane (see also \citet{inn82,wk88}). This holds true along both
major axes, but seems particularly apparent in the north-east.

An interesting high velocity HI cloud lying near the northern minor
axis of M31 was discovered by \citet{dav75}.  The feature lies
approximately 1.5\arcdeg\ ($\approx 20$~kpc) north-west of the center
of M31 at (0.9\arcdeg W, 1.1\arcdeg N) , has an extent of
$\sim1$\arcdeg\ and a heliocentric velocity of $-450$~km/s (see Figure
\ref{sats}).  It lies just above NGC~205 and, rather curiously, along
the projected extension of the giant stellar stream discovered in the
southern half of M31.  Based on the velocity of the cloud, it would
appear unlikely that it has any connection to the north-west disk, nor
for that matter to either M32 (V$_{helio}=-197$~km/s) or NGC~205
(V$_{helio}=-242$~km/s).  The cloud has a peak column density of
$4\times10^{19}$~atom~cm$^{-2}$, a gaseous velocity dispersion of
9~km/s and, if at the distance of M31, a gas mass of
$5\times10^{6}$~\msun; it thus represents a fairly typical compact high
velocity cloud.  Inspection of CMDs and detailed spatial density and
metallicity maps centered on the cloud position reveal nothing unusual
at this location.  An alternative explanation for the cloud is that it
is very local and related to the Magellanic Stream \citep{dav75}; the
non-detection of an associated stellar counterpart to the cloud may be
more consistent with this interpretation.

\citet{newt77} also discovered an apparently detached HI cloud at large
radius along the north-eastern major axis of M31.  Located at a
projected distance of $\sim 3\arcdeg$ or $\sim 34$~kpc, this cloud has
a velocity which differs by 100~km/s from that expected for local  disk
rotation.  The global properties of the cloud are similar to those
inferred for the Davies cloud.  The cloud lies at (1.5\arcdeg E,
2\arcdeg N) from the center of M31 and near our current survey limit in
this quadrant; our maps reveal no obvious spatial substructure here,
but the mean metallicity of the RGB stars is somewhat above average
(see Figure \ref{metal}).  The HI feature may be a signature of a major
disk disturbance in the north-east quadrant, possibly related to the
northern spur feature which is located roughly 0.5\arcdeg\ further in.

\subsection{The Origin of the Substructure}

Substructure in the form of accreted and/or disrupted satellites is a
generic prediction of hierarchical galaxy assembly.  Even if the
satellites are accreted before significant star formation has occurred,
they will still interact gravitationally with the stellar disk of the
host, causing tidal heating, distortion and possibly even warping.  
Our WFC survey has led to the discovery of significant spatial and
chemical substructure in the outskirts of M31, the most prominent of
which is characterised by Figure \ref{cartoon}.  We first address
whether any of this substructure is directly associated with the
luminous satellite companions of M31.

M31 is known to possess approximately 15 satellites, of which the dwarf
ellipticals (dE) M32 and NGC~205 are the most proximate, lying at
projected distances of 5 and 9~kpc respectively, and among the most
luminous \citep{mateo98b}. Figure \ref{sats} illustrates where the
closest satellite companions lie in projection with respect to M31.
Apart from the $`$clump' of stars close to the luminous cluster G1 and
the northern spur, the nature of which are not yet clear,  our survey
has failed to reveal any obvious (\ie still intact) additional
satellites out to $\sim 3\arcdeg$ from the center of M31.  For
reference, all known Milky Way satellites (including the Sagittarius
dwarf) have surface brightnesses and angular extents that would have
rendered them easily detectable in our survey, if lying within a
projected radius of $\approx55$~kpc of M31.

  M32 and NGC~205 are known  to exhibit anomolous surface brightness
profiles and/or outer isophote twists suggestive of tidal interaction
and disruption (\eg \citet{hodge73,choi02}, see also Figure
\ref{distort}).  In addition, both display a variety of puzzling
properties for galaxies of the dE class. M32 exhibits a strong
intermediate-age stellar component,  in addition to a classical old
stellar component,  and shows evidence for a large metallicity spread
with a mean just below solar \citep{grillmair96,davidge00}. The
signatures of an extended and complex star formation epoch are even
more obvious in NGC~205, which possesses cold gas (both atomic and
molecular) and dust as well as young stars \citep{lee96,young97}.
Furthermore, the HI shows a well-defined velocity gradient (unlike the
stars) and is confined to the very inner regions of the optical galaxy;
this inconsistency between gas and stars suggests that the gas may have
been recently captured, perhaps due to a passage through the disk of
M31.   The geometrical alignment of the two satellites with the
southern stream, and the broadly similar metal abundances initially
suggested  that either or both of them could be responsible for the
tidal feature \citep{ibata01b}.  With a significant fraction of the
northern half of M31's halo now mapped, we have failed to identify any
obvious continuation of the stream on the northern side of NGC~205.
Given that tidal interactions give rise to both leading and trailing
streams from the disrupted satellite, such a feature would be expected
if NGC~205 were the origin of the stream.  M32 or a third system
(either more distant, or now completely-cannibalized?) now appear the
most promising candidates for the origin of the stream.  We have
previously remarked on the possible connection between the stream and
the similarily metal-rich, but lower-level, stellar overdensity near
the north-eastern side of the disk (see Figure \ref{cartoon}).  If this
association is correct, it implies  the projected orbit of the
disrupting satellite wraps tightly around the M31 nucleus, and does not
cross into the northern half of the halo of M31.  On the other hand,
the lack of a northern counterpart to the stream might simply reflect
the fact that the current orbit of the satellite has yet to traverse
into this part of the halo, or else that the stream in this region has
already dispersed and is no longer visible as a spatial or chemical
density enhancement.

Other satellites of possible relevance for the interpretation of the
M31 substructure are the dwarf Spheroidal systems, Andromeda~I and
Andromeda~III.   Both lie just beyond the boundary of our current
survey to the south-west of M31, consistent with the general direction
of the  stream. The stellar populations in these systems are
predominantly metal-poor however, a fact that  would seem to argue
against them being the origin of the more metal-rich stream.
Specifically, And~I and And~III have mean RGB metallicities of $[$Fe/H$]
\sim -1.5$ and  $-1.9$ respectively \citep{dacosta96,dacosta02}
compared to the value of $[$Fe/H$] \gtrsim -0.7$ inferred for the
stream.  M31's largest satellite, M33, lies at a projected distance of
$\sim 15\arcdeg$ towards the south-east. There is no evidence at
present to connect M33 with any of the substructure seen around M31,
however the potential role this system could play within the M31
subsystem should be borne in mind.
   
The extreme compactness of M32, coupled with the very high central
surface brightness and the dwarf-like luminosity, has led many authors
to suggest that M32 is the leftover core of a larger galaxy, evolved to
its current state as a result of prolonged tidal $`$harrassment$'$
within M31's halo (\eg \citet{faber73,bekki01,graham02}).  Such a
hypothesis may also explain the puzzling observation that M32 has no
globular clusters; a system this luminous would be expected to possess
at least $\sim 15-20$ of them \citep{harr91}.  The broad agreement
between the stellar metallicity and dispersion in M32 and in the M31
halo \citep{grillmair96,durrell01} leads one to speculate whether the
former could be the origin of not only the stellar stream, but also
much of the field halo.  If the M31 halo was significantly polluted by
stars from a fairly massive companion, this could also account for the
higher density/size of the stellar halo \citep{reitzel98} compared with
that of the Milky Way.  For such a scenario to be viable, M32 must have
already made several revolutions around M31 in order to account for the
high covering factor of the metal-rich population. Dynamical friction
due to stars and dark matter in M31's halo will cause the orbit of M32
to spiral inwards with time, while any non-sphericity in the potential
and/or close passages to the disk will cause the orbit to precess.  The
significant velocity dispersion of M32 stars \citep{vdm94} implies that
once pulled off, these stars will merge rather quickly into the halo
population and only the most recently stripped stars will still appear
as a coherent structure.  \citet{pen02} have shown that tidal debris
from a disrupting satellite will spread out even faster if the host
halo is significantly flattened, with a timescale of 3--4~Gyr before
total disruption of the companion.  Detailed dynamical modelling of 
satellite orbits within the potential of M31 will be presented in a future paper.

Few constraints exist on the origin of the major-axis substructure at
present.  It would appear unlikely that the G1 clump and the northern
spur are related to each other (for example, resulting from an object
orbiting close to the plane of the disk) due to the different mean
colors of the RGB stars in these regions and the presence of a
prominent intermediate-age component in the spur but not in the clump.
The G1 clump is particularly intriguing given its proximity to the
anomolous globular cluster G1.  G1 is among the most luminous and most
massive globular clusters in M31 and is rather unique in having an
intrinsic metallicity spread, suggestive of self-enrichment
\citep{meylan01}.  The outer isophotes of the cluster are distinctly
elongated, and the direction of this elongation appears to be in the
same sense as that of the G1 clump.  Meylan \etal have argued that G1
might be the core of a disrupted dE and such a scenario is supported by
the fact that the metallicity reported for G1 ([Fe/H]$=-0.95$) agrees
quite well with that inferred for the stellar overdensity, \ie slightly
below that of the average halo [Fe/H]$\sim -0.7$.  On the other hand,
the significant spatial offset (0.5\arcdeg\ or 7~kpc if at the distance
of M31) seen between G1 and the peak of the stellar overdensity is
difficult to understand in this picture, as is the fact that the
luminosity and mean metallicity of the clump place it well off the
luminosity-metallicity relation defined by Local Group dE and dSph
galaxies (\eg \citet{mateo98b}).  Given the metallicity
of the G1 clump as determined from mean RGB colour, the
luminosity-metallicity relation predicts a luminosity more than two
magnitudes brighter than that observed.  An alternative explanation for
the feature is that it is a highly warped section of the far outer
disk, perhaps perturbed or $`$torn off' by a previous tidal
interaction.  The clump lies just beyond the outermost HI contour of
\citet{newt77} which warps southward in this quadrant, compared to the
northern offset displayed by the clump.  If the disk interpretation is
correct, the fact that the gas and stars currently exhibit different
behaviours here could imply the interaction happened a long time ago.
Future photometric and kinematic observations will aid significantly in
discriminating between these two possibilities.

The location of the northern spur near the north-eastern major axis of
M31 and in the same direction of the gaseous warp provides strong
support for the association of this  feature with a severe warp in the
outer stellar disk.   This interpretation would naturally explain the
metal-rich nature of the population in the spur, as well as the
presence of a young-to-intermediate age component, as these are
properties which are also known to characterise stars in the outer
parts of the disk \citep{morris94,ferg01}).  \citet{wk88} noted that if
the northern spur was an extension of the disk, it would be the most
extreme example of a warped stellar disk ever found.  It would also be
highly asymmetric.   Tidal forces due to an infalling satellite could
possibly excite such a strong warp \citep{binn92,huang97}, however the
identity and location of the putative perturber is not obvious at
present.

\section{Summary}

Our INT WFC survey currently maps an area of $\approx 25~\sq\arcdeg$
around M31, extending to a semi-major axis of 55~kpc, and allows the
first uninterrupted study of the density and color distribution of
individual red giant stars across a large fraction of an external disk
galaxy.  We have found evidence for both spatial and metallicity (as
inferred from color information) substructure, which are often, but not
always, correlated. In addition to the giant stellar stream reported by
\citet{ibata01b}, the data reveal the presence of significant stellar
overdensities at large radii close to the south-western major axis, in
the proximity of the very luminous globular cluster G1, and near the
north-eastern major axis, coinciding with and extending beyond the
$`$northern spur'.   The most prominent metallicity variations are
found in the southern half of the halo, where two  structures with
above average metallicities are apparent.  One of these coincides with
the aforementioned giant stellar stream, whereas the other corresponds
with a much lower-level stellar overdensity.

Our results contrast with, but do not conflict with, the findings of
previous M31 stellar population studies that have found a rather
homogeneous population in the halo and outer disk. These studies have
mostly  focused on inner halo fields or fields at large radius along
the southern minor axis, neither of which are found to exhibit
significant substructure within the limits of our current survey.   Of
the newly-detected substructure, only the northern spur appears to have
a counterpart in published HI maps \citep{newt77}.  Further, our data
reveal no spatial or chemical overdensity coincident with the high
velocity cloud discovered by \citet{dav75} which lies just north of the
satellite companion NGC~205, along the projected northern extension of
the giant stellar stream.

We have argued that the stellar  substructure seen in the outskirts of
M31 is likely due to a variety of past and ongoing interaction and
accretion events.  The lack of any obvious northern counterpart to the southern
stellar stream suggests that either the disrupting satellite has yet to
traverse into the northern half of the halo, or that the projected
orbit wraps tightly around the center of M31.  Indeed, there is
tentative evidence to connect the stellar stream to the stellar and
chemical enhancement seen towards the north end of the disk.   Several
pieces of evidence suggest that M32 could be the origin of the stream;
these include the geometrical alignment of the two objects, the
similarity in the mean metallicity and dispersion of their constituent
stars and the wide range of peculiar properties exhibited by M32, all
of which are consistent with it having  once been a considerably more
massive and luminous galaxy.   One may even speculate that stars
stripped from M32 have polluted a large fraction of the M31 halo,
providing a viable explanation for the long-standing puzzle of why
M31's field halo is more metal-rich and  denser/larger than that of the
Milky Way.  Alternatively, the stream may be due to another satellite,
perhaps now completely cannabilized or else lying beyond our current
survey limits.

The stellar substructure located close to the major axes -- the
northern spur and the G1 clump -- could plausibly result from tidal
distortion and disruption of the outer disk due to, for example, the
close passage of a massive satellite.  If this explanation is correct, the
interaction  responsible for producing the G1 clump must have occurred
some time ago, given the lack of evidence for a strong intermediate-age
component and the slightly lower than average metallicity in this
structure.  On the other hand, the remarkable proximity and overall
alignment between the clump and very luminous globular cluster, G1,
 supports the alternative hypothesis that the two have a common origin.
\citet{meylan01} have proposed that G1 may be the core of a stripped
dwarf elliptical galaxy; this hypothesis clearly warrants further
consideration.  However, a satisfactory model must also account for the
$\sim 0.5\arcdeg$ projected offset seen between the peak of the stellar
overdensity and the globular cluster.

The Milky Way and M31 are the only two massive galaxies for which we
have detailed information about how the properties of individual stars
vary across a large fraction of the halo.   The mean metallicity and
size/density of the stellar halos in these systems differ substantially
and this has sometimes been cited as evidence for different formation
mechanisms (\eg \citep{durrell01}).  Our INT WFC survey reveals that,
like the Milky Way,  the stellar populations at large radii in M31
exhibit significant substructure.  This substructure is expected in
hierarchical CDM models of galaxy formation, either as a direct result
of disrupting satellites or as a by-product of tidal interactions
between infalling satellites and the fragile outer disk.  The fact that
luminous, and clearly distorted, satellites lie close to the center of
M31 suggests that the stellar halo differences between that galaxy and
the Milky Way may be the result of a more active accretion/merger
history.   Essential to the interpretation of our newly-detected
substructure will be deep photometric and spectroscopic observations of
stars in the outskirts of M31. An approved Cycle 11 HST program with
the newly-installed Advanced Camera for Surveys is aimed at obtaining
deep color-magnitude diagrams, reaching several magnitudes below the
horizontal branch, in several fields where we have detected halo
substructure, and in two locations in the far outer disk.  These data
will provide  more quantitative information about the ages and
metallicities of stars in the various substructures and will help to
constrain the accretion/merger history of the galaxy.    Spectroscopic
observations with 8-m telescopes will be essential for providing
kinematical constraints to be used in modelling the satellite
interactions, and for probing the shape and potential of M31's massive
dark halo.

Finally, it is worth noting that M31 is being used as a target in
several large experiments that search for microlensing events due to
Massive Compact Halo Objects (MACHOs).  Without further detailed
modelling, it is impossible to determine where the giant stellar stream
lies with respect to center of M31.  If it lies in front of the galaxy,
it will produce a population of lenses that are more probable than the
average M31 bulge star to lens M31 bulge stars, whereas if it lies
behind the galaxy, it will produce an extra population of sources which
are more likely to be lensed by M31 bulge stars.  A careful appraisal
of the stream, as well as other M31 stellar substructure, will
therefore be required in order to interpret the microlensing event
rates towards M31.

\acknowledgments

We thank the Institute of Astronomy, Cambridge and the Anglo-Australian
Observatory for hospitality during various collaborative visits.

\clearpage

\clearpage

\begin{figure}
\figurenum{2}
\caption{(a)~A standard coordinate projection of the surface density of
blue RGB stars across our current $\approx 25 \sq \arcdeg$ survey
area.  The inner and outer ellipses are drawn assuming a position angle
of 38.1\arcdeg, derived from analysis of the light distribution on a
scanned Palomar Sky Survey plate.  The outer ellipse denotes a
flattened ellipsoid (aspect ratio 3:5) of semi-major axis length 55~kpc
and indicates the current spatial extent of the survey. The inner
ellipse has a semi-major axis of 2\arcdeg ($\approx 27$~kpc) and
represents an inclined disk with $i=77.5$; the optical disk of M31 lies
well within this boundary.   The few white blotches indicate regions
contaminated by saturated stars.  The dwarf companions M32 and NGC~205
lie at (0\arcdeg,$-0.4$\arcdeg) and ($-0.5$\arcdeg,$0.4$\arcdeg)
respectively.  Much substructure is seen at large radii, including the
giant stellar stream and stellar overdensities at both extremes of the
major axis.  No corrections have been made for foreground or background
contamination.}
\end{figure}

\clearpage

\begin{figure}
\figurenum{2}
\caption{(b)~Same as (a)
except showing the surface density of red RGB stars. Note the lower
Galactic foreground contamination on this map. Comparison with (a) clearly indicates that the morphology of the substructure
varies as a function of color.  \label{rgb}}
\end{figure}

\clearpage

\begin{figure}
\figurenum{3}
\caption{A standard coordinate projection of the surface density of stars
lying redward of and above the RGB tip. These stars are likely to be
intermediate-age, moderate metallicity thermally-pulsing asymptotic
giant branch stars. Of the substructure visible in the RGB map, only
the northern spur is clearly detected on the AGB map; there is also
marginal enhancement in the vicinity of the G1 clump.  No corrections
have been made for foreground or background  contamination. \label{agb}}
\end{figure}

\clearpage

\begin{figure}
\figurenum{4}
\caption{Examples of two color-magnitude diagrams from our WFC survey.
The left panel represents a field in the giant stellar stream and, the
right panel, a field in the G1 clump.  Over-plotted are the same
globular cluster fiducial sequences as shown in Figure \ref{sel}.  Most
of the stars more luminous than the RGB tip are foreground
contaminants.  Stellar density varies between the CMDs as a result of
the different overdensities of the features.  We argue the different
mean colors and widths of the RGB primarily reflect intrinsic
metallicity variations.
\label{cmds}}
\end{figure}

\clearpage

\begin{figure}
\figurenum{5}
\caption{A map of the variation in 
mean RGB color across M31 (see text for details).   Each WFC pointing
is represented as a color-coded polygon.  The $`$average' halo RGB
color is denoted as green; assuming an old stellar population, this
corresponds to a metallicity of [Fe/H]$\sim -0.7$, \ie similar to
47~Tuc. Yellow, orange and red regions indicate progressively redder
mean colors, and hence higher metallicities.  Turquoise and light blue
regions indicate progressively bluer mean colors, and hence lower
metallicities. The full metallicity range spanned by the plot (red
through light blue) is $\approx 0.7$ dex. Dark blue regions represent
pointings where the mean stellar density was too low to make a
definitive measurement of stellar color.    \label{metal}}
\end{figure}

\clearpage

\begin{figure}
\figurenum{6}
\centerline{\psfig{file=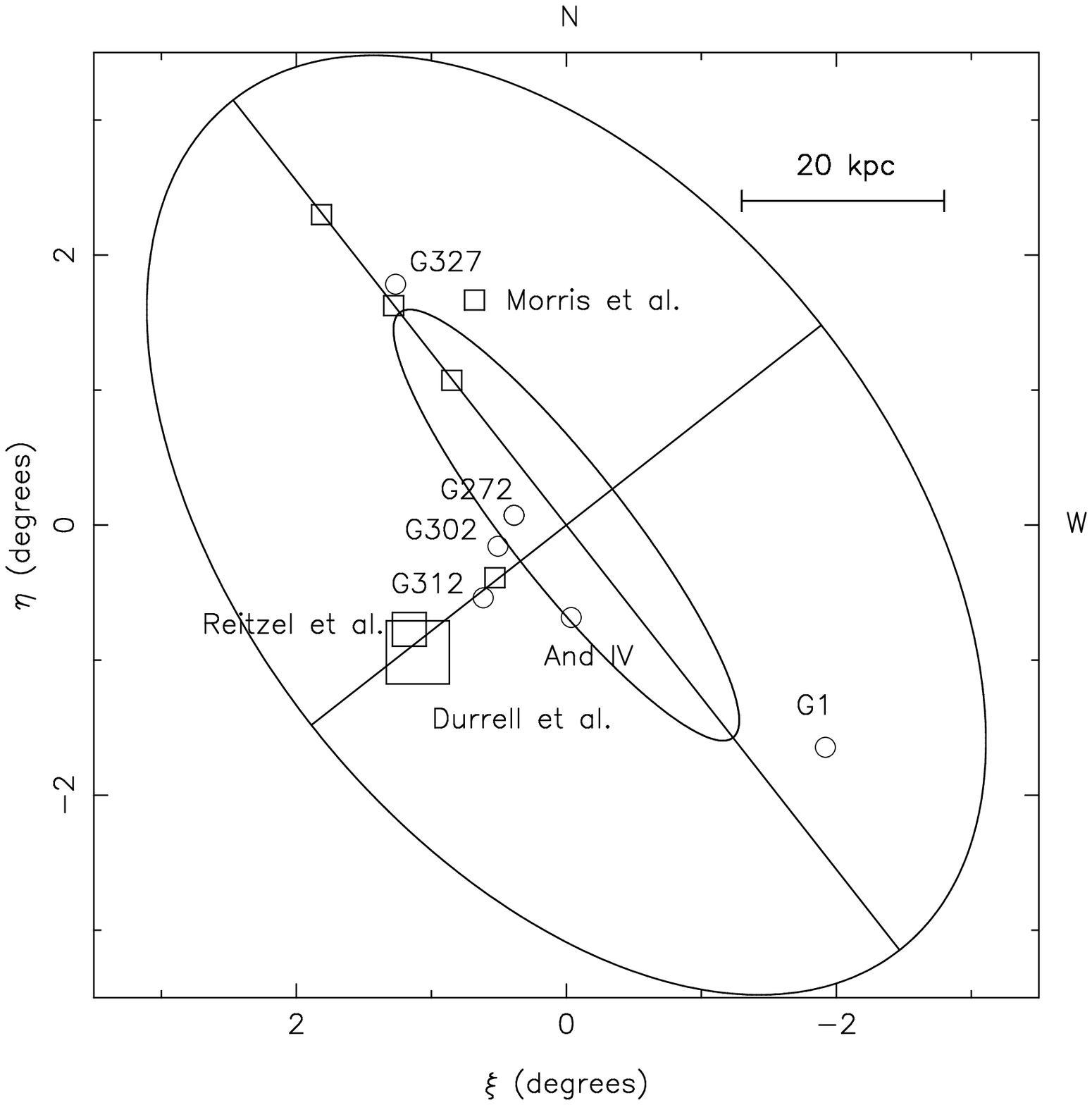,angle=-90,height=15cm}}
\caption{A diagram showing the locations of  field population studies
 in the outskirts of M31. The squares show fields targeted by the
ground-based studies of \citet{morris94,reitzel98} and
\citet{durrell01}.  The circles represent the HST/WFPC2 studies of
\citet{holland96,rich96,sara01,ferg00} and \citet{ferg01}.  Only the Morris
\etal $`$spur' field and the Rich \etal G1 field lie close to halo
substructure discovered in the present survey.\label{prev}}
\end{figure}

\clearpage

\begin{figure}
\figurenum{7}
\centerline{\psfig{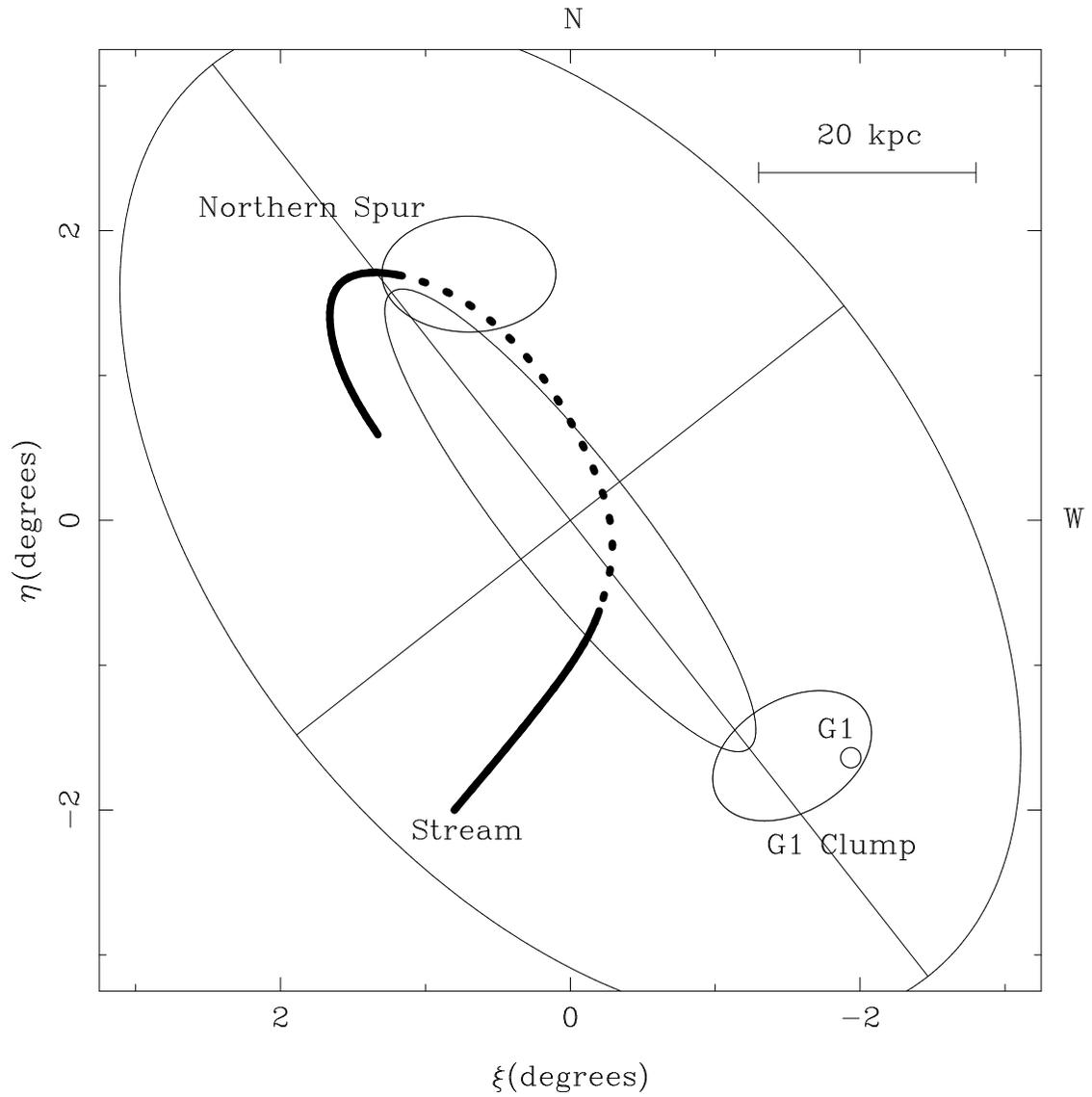}}
\caption{A cartoon illustrating the most prominent spatial and chemical
substructure discovered in our WFC survey.  A possible projected orbit
of the giant stellar stream is indicated; the stream may connect to the
enhancement seen in  the spatial and chemical maps of the northern half
of the disk.  The contours delineate the approximate extent and
orientation of the northern spur and G1 clump; the location of the G1
globular cluster is also indicated.\label{cartoon}}
\end{figure}

\clearpage

\begin{figure}
\figurenum{8}
\centerline{\psfig{file=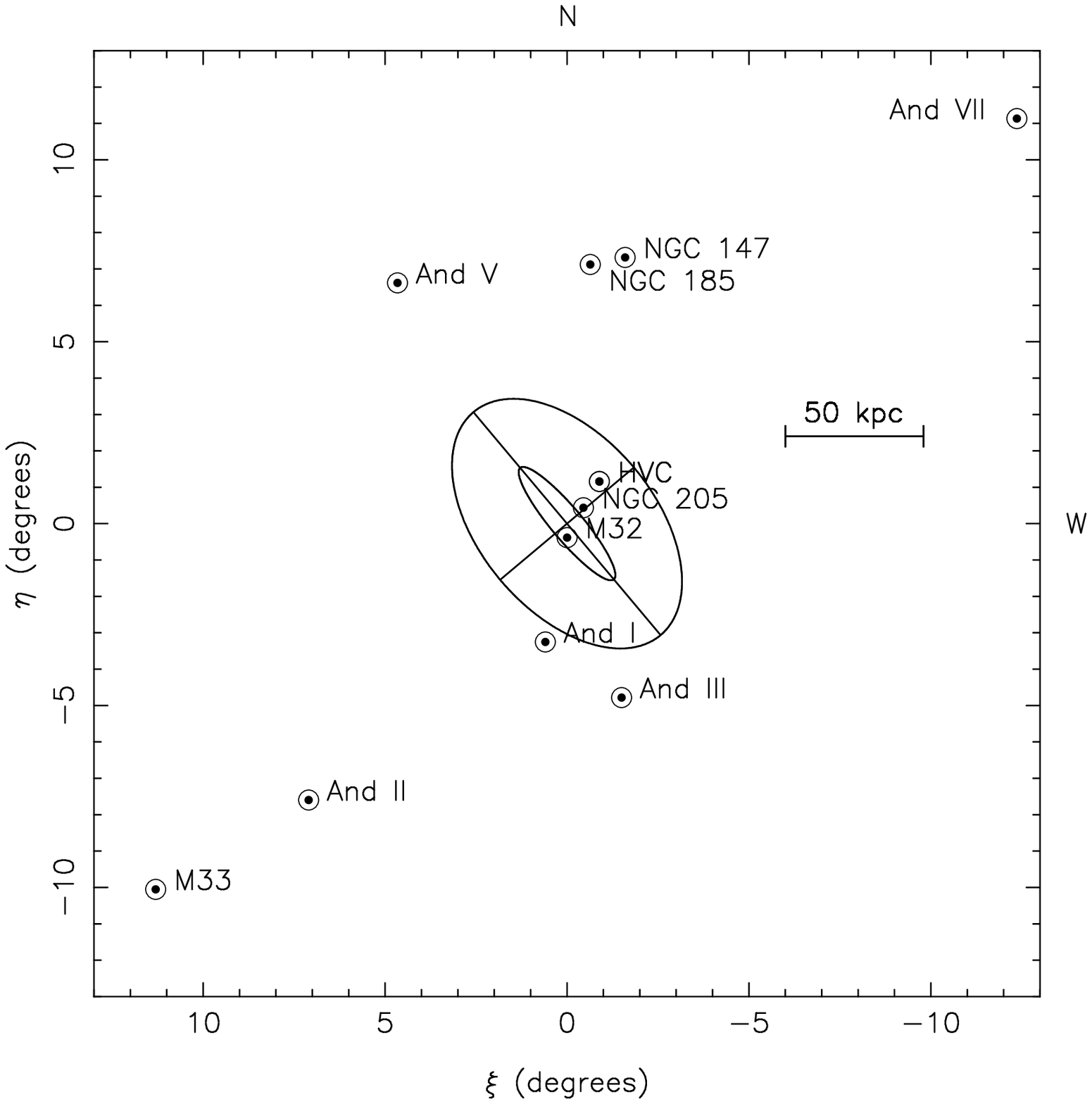,angle=-90,height=15cm}}
\caption{The projected distribution of the closest satellite companions around
M31. Also shown is M31's nearest massive companion, M33 and the position of
the neutral hydrogen cloud detected by \citet{dav75}. As
 previously, the outer ellipse denotes the approximate  limit of our
current WFC survey.\label{sats}}
\end{figure}

\clearpage

\begin{figure}
\figurenum{9}
\caption{Isopleth maps of M31 from an APM scan of a 75 min exposure
Palomar Schmidt IIIaJ plate taken by Sydney van den Bergh in 1970.  The
lowest isophote contoured is at a level equivalent to B$=$27 mag
arcsec$^{-1}$.   The right-hand panels show the distorted outer
isophotes of M32 and NGC~205 after substracting an
elliptically-averaged M31 profile.  \citet{cepa} quote tidal radii for
M32 and NGC~205 of 0.84 kpc ($0.06^\circ$) and 1.93 kpc ($0.14^\circ$)
respectively, which corresponds to the relatively unperturbed inner
parts of the profiles.  The location of the  \citet{dav75} HI cloud
is also indicated on the main figure.  In the southern half of the
galaxy, the M31 outer isophotes are distorted in the direction of the
stream.  Although the lowest contour level is at the limit of
relibability of the photographic data, this particular feature is
present in other digitised wide area photographic data (\eg
\citet{wk88}).\label{distort}}
\end{figure}

\end{document}